# Field/source duality in topological field theories

**D. H. Delphenich** [*]

Physics Department, Bethany College, Lindsborg, KS USA 67456

**Key words**  Field singularities, topological electromagnetism, topological field theories

**PACS** 02.40.Re, 03.50.De. 11.10-z

The relationship between the sources of physical fields and the fields themselves is investigated with regard to the coupling of topological information between them.  A class of field theories that we call *topological* field theories is defined such that both the field and its source represent de Rham cocycles in varying dimensions over complementary subspaces and the coupling of one to the other is by way of an isomorphism of the those cohomology spaces, which we refer to as *field/source duality*.  The deeper basis for such an isomorphism is investigated and the process is described for various elementary physical examples of topological field theories.

### Contents

---

[*] E-mail:  delphenichd@bethanylb.edu

## 1 Introduction

One of the most fundamental issues in physical field theories is how the information associated with the field sources is coupled to the information in the field itself. For instance, in the inverse-square case that describes either the field of a static point charge or the gravitational field of a static point mass, in the Newtonian approximation, this coupling takes the form of associating a scalar – charge or mass, resp. – with the source point and a vector field or 1-form at every other point of the space whose magnitude is proportional to that scalar, along with purely geometrical information that describes how the field varies with distance from the source and the direction that it points in.

A topologically significant issue regarding fields is whether the points that represent the field source are points at which the field itself is actually defined or if the set of these points should be properly the complement of the points at which the field is defined. For instance, this is necessarily the case with an inverse-square field, which has a pole at the source point.

To be sure, in physics, the source of a field does not always have to represent a singularity. For instance, one sometimes regards a field as being "driven by its boundary conditions," such as when one solves for a potential function in a region by specifying its value on the boundary of that region. Similarly, a disturbance in a medium that initiates the propagation of a wave does not always have to be topologically distinct from the rest of the medium, such as the motion of the hand that sends a pulse down a bullwhip. Contrast this with the case of a stone dropped into a pond, which momentarily creates a hole in the surface, and the hole can be regarded as either an example of a non-trivial homology class in dimension one that serves as the source of the wave field or a boundary on which to define initial conditions (i.e., Cauchy data).

Hence, we see that the scope of our present discussion does not encompass all possible field/source relationships, but simply an important class that presents itself at the most elementary level. A further restriction in scope is due to the fact that the fields that we will be considering often take the form of "external" field solutions, as opposed to the "internal" field solutions, which are defined on the source points; by this, we generally mean the points of the support of the source distribution.

We then regard the space or spacetime manifold $M$ as being decomposed into two types of points: a closed subset $S$ of "singular points" at which the source distribution is defined and its open complement $G$ of "generic points" at which the field is defined. Furthermore, we shall assume that $S$ represents the set of boundary points of $G$; i.e., limit points that are not generic. This is simply to suggest that source points are not topologically isolated from the fields that they generate. In order to justify the use of the term "duality" in this situation as being indicative of some sort of fundamental isomorphism, we then concentrate on a class of field theories in which one can find the same essential information contained in the source as in the field itself.

We shall call a field theory a *topological field theory* [1] when the source points define a homology class, in some choice of homology, and the field points define a cohomology class in dimension $k$, in perhaps some other sense. Furthermore, the source points are associated with a source distribution, such as a charge or current density, which we then regard as a cohomology class over $S$ in some dimension $m$. Hence, we shall say that a topological field theory exhibits *field/source duality* when there is an isomorphism of. $H^k(G; R)$ and $H^m(S; R)$.

Of course, to the pure mathematician this isomorphism would suggest that there is something redundant about describing both the cohomology of $S$ and that of $G$. They are more inclined to simply regard the fact that the cohomology of $G$ is non-vanishing in dimension $k$ as the "source" of the field. The reason that it is still physically non-trivial to discuss both cohomologies independently is that ultimately the nature of $S$ is somewhat enigmatic in the eyes of experimental physics. Hence, any attempt to mathematically model the topology of $S$ is limited by the state of the art in phenomenology.

The homologies that we shall use in this study are the homology of differentiable singular cubic chains with real coefficients and the de Rham homology of $k$-vector fields on an orientable manifold, which is given a divergence operator that is Poincaré-dual to the exterior derivative. The corresponding cohomology will be the de Rham cohomology, which is defined by exterior differential forms and the exterior derivative operator.

Of course, the prototype for this sort of field theory is electromagnetism, in which the field is a closed 2-form $F$ and the source is a conserved – i.e., divergenceless – vector field $\mathbf{J}$. When one is given a linear electromagnetic constitutive law $\kappa: \Lambda^2(M) \to \Lambda_2(M)$ that associates an excitation bivector field $\mathsf{h}$ to the field strength 2-form $F$ the other Maxwell equation $\delta\mathsf{h}(F) = \mathbf{J}$ then the operator $\delta\kappa: \Lambda^2(M) \to \Lambda_1(M)$ maps a de Rham cohomology class in dimension two, namely $[F]$, to a de Rham homology class in dimension one, namely $[\mathbf{J}]$.

Although one is taught to think of this association as coming from the various forms of Stokes's theorem, such as Gauss's law in electrostatics and Ampère's law in magnetostatics, it is interesting that both of those laws are actually *inapplicable* in the most elementary cases of point charges and line currents, since the fields diverge at those points. Hence, the fields are defined only on the complements of points in $R^3$ or lines in $R^3$, and these complementary topological spaces have non-trivial cohomology in dimensions two and one, respectively. Thus, one finds that although the total flux of the electric field strength $\mathbf{E}$ through any sphere $S_r^2$ of radius $r$ and centered on a point charge $Q$ at the origin is $Q$, just one expects, nevertheless, that the divergence of $\mathbf{E}$ vanishes everywhere in the interior of the ball $B_r^3$ that $S_r^2$ bounds, except at the origin, where it is not defined. Hence, since $S_r^2$ is a 2-cycle, but the punctured ball $B_r^3 - \{0\}$ that it bounds is not even a 3-chain in $R^3 - \{0\}$, Gauss's law is inapplicable to the situation, and the only way that one can associate the total flux of $\mathbf{E}$ through $S_r^2$ with $Q$ is to make that a basic *postulate* of the field theory. This amounts to associating the essential information in the homology class defined by the 0-cocycle $Q\delta^0$ with the cohomology class defined by the 2-cocycle #$\mathbf{E}$. (In this expression, the 0-cocycle $\delta^0 \in H^0(\mathsf{S}; \mathsf{R})$ is a generator that takes the 0-cycle $\{0\}$ to 1 and all other 0-cycles to 0 and #: $\Lambda_*(M) \to \Lambda^*(M)$ is the Poincaré duality isomorphism).

---

[1] We should probably say topological *classical* field theories, to distinguish them from the topological *quantum* field theories of Witten, Atiyah, et al., but the author believes that if the scientists of antiquity were not actually considering any such modern refinements as cohomology and fiber bundles then calling such theories "classical" is unnecessarily pejorative, as it suggests anachronism; perhaps "neo-classical" would be a more appropriate descriptive.

We point out that there is a subtle chain of associations in the above example between the surface integral of the 2-form #$\mathbf{E}$ over $S_r^2$, the integral of $\delta\mathbf{E}$ over the volume $B_r^3 - \{0\}$ (which is, a 3-chain in $\mathsf{R}^3$, but not in $\mathsf{R}^3 - \{0\}$), and the integral of the charge density 3-form #$\rho = Q\delta_0\varepsilon$ over the ball $B_r^3$ in $\mathsf{R}^3$ that $S_r^2$ bounds, where $\varepsilon$ is the volume element on $\mathsf{R}^3$. Apparently, the weak link is in associating the volume integral of $\delta\mathbf{E}$ over $B_r^3 - \{0\}$ with the volume integral of #$\rho$ over $B_r^3$ when Gauss's law was already shown to be inapplicable at the first step. We can express these three integrals concisely in terms of the basic pairing of $k$-cochains and $k$-chains by integration as $<\#\mathbf{E}, z_2>$, $<\#\delta\mathbf{E}, c_3>$, and $<\#\rho, c_3>$. Since we see that these three expressions are connected by equalities only when the 2-cycle $z_2 = \partial c_3$, we also see that such an association must be introduced as a basic assumption.

Naturally, there is something both topologically and physically unsatisfying about anything that must be assumed rather than derived, so we shall investigate the extent to which the association of the information that $<\#\mathbf{E}, z_2>$ represents with the information that $<\#\rho, c_3>$ represents can be derived from more fundamental assumptions. For instance, one notes that since $Q$ is concentrated at a point, the 3-chain can be homotopically contracted to a 0-chain $\{0\}$, the 3-cocycle #$\rho$ to a 0-cocycle $Q\delta^0$ and the information in $<\#\rho, c_3>$ is the same as in $<Q\delta^0, \{0\}>$. More generally, one often applies deformation retractions to reduced tubes to curves and slabs to planes. Furthermore, in order to make actual homology classes out of the source points, one must often reduce the curves and planes to points, at least when the curve is not a loop. Hence, we shall discuss the notion of field theories that are topologically equivalent to topological field theories.

## 2  Topological preliminaries

Although the basic concepts of homology theory (see [1-4]) are assumed, some attempt is made to make the presentation more self-contained, so these concepts are also defined for the sake of discussion below.

We begin by briefly summarizing the basic concepts of the real homology of differentiable singular cubic chains.

### 2.1  Differentiable singular cubic homology

Although the basic building blocks of differentiable singular cubic homology are $k$-dimensional cubes $I^k$ in $\mathsf{R}^n$, and are therefore not differentiable manifolds, strictly speaking, this is not a serious limitation since differentiation is a local operator, so no matter how one chooses to extend a function $\sigma_k : I^k \to M$ to a differentiable function on an open neighborhood of $I^k$, the restriction of $d\sigma_k$ to $T(I^k)$ will remain the same. Hence, this is the sense of the word "differentiable" we intend by defining a *differentiable singular cubic k-simplex* to be such a map. A *differentiable singular cubic k-chain* with coefficients in a ring $R$ is then a finite formal sum [2] of the form:

$$c_k = \sum_{i=1}^{N} \alpha_i (\sigma_k)_i \,, \qquad \alpha_i \in R \,. \tag{2.1}$$

The most natural choice of ring is $\mathsf{Z}$, the integers, since the multiplication by a integer evokes an intuitive picture of the repetition of a simplex in the sum, but eventually we shall be primarily concerned with the ring (field, really) of real numbers $\mathsf{R}$. One can then think of the coefficient of a simplex as essentially a "total charge" that is associated with the object.

The simplest non-trivial coefficient ring is $\mathsf{Z}_2 = \{0, 1\}$, which plays an important role in the topology of orientability and the Stiefel-Whitney characteristic classes. In an expression for a $\mathsf{Z}_2$ $k$-chain of the form (2.1), a basic simplex would either appear with a + sign for its coefficient or not at all. This is to be distinguished from $k$-chains – in particular, $k$-cycles, which we discuss below – that can have any integer for a coefficient, but some of them vanish when multiplied by some integer $p$; these are referred to as $k$-cycles with *torsion*. Fortunately, such complexities will be irrelevant to most of what follows since we will be using real coefficients, which allow for no torsion factors in homology or cohomology.

For the sake of brevity, we shall henceforth generally abbreviate the phrase "differentiable singular cubic chain" to simply "chain" when no confusion will arise.

It is important to point out that the inclusion of the word "singular" is not casual, since the maps of the $k$-cubes into $M$ are not required to be embeddings. Hence, one should keep in mind that the dimension of the image could very well be less than $k$; indeed, a singular $k$-simplex can take $I^k$ to just one point. (This is useful to keep in mind when one is considering homotopies of chains.)

We shall call a chain *homogeneous* if all of the cells or simplexes ([3]) are of the same dimension. More generally, an inhomogeneous chain can always be partitioned into homogeneous subchains. Hence, if we denote the $R$-module of all differentiable singular cubic chains in $M$ with coefficients in $R$ by $C_*(M; R)$ then we see that this $R$-module is expressible as a direct sum of homogeneous sub-modules:

$$C_*(M; R) = C_0(M; R) \oplus C_1(M; R) \oplus \ldots \oplus C_n(M; R) \,. \tag{2.2}$$

---

[2] For those who find formal sums vague and non-rigorous, it should suffice to say that one can actually make this expression more rigorous, but it would distract from the immediate argument to do so.

[3] The author's justification for using this plural of the word "simplex," as opposed to "simplices" is that "simplexes" is more consistent with the plural of "complex"; i.e., the plural of "sim(ple com)plex." One will note that this is the plural that was originally used by Eilenberg, Steenrod, Hilton, Wylie, and others. Sometime back in the Sixties, the researchers in homologies seemed to have had second thoughts and switched to "simplices," apparently by analogy with "vertices."

This $R$-module is a real vector space in the case where the coefficient ring is $\mathbb{R}$, and is generally infinite-dimensional, as long as $M$ is not a finite point set.

When $k > 0$, we define the $i^{th}$ *zero face* of $I^k$ to be the $k-1$-cube defined by $\phi_{i0} = (\dots, 0, \dots)$, with a 0 in the $i^{th}$ coordinate, and the $i^{th}$ *unit face* to be $\phi_{i1} = (\dots, 1, \dots)$. We define the *boundary* of $I^k$ to be the formal sum of its oriented $k-1$-faces:

$$\partial I^k = \sum_i (\phi_{i1} - \phi_{i0}) . \tag{2.3}$$

One then defines the boundary of a $k$-simplex $\sigma_k$ so that it equals the $(k-1)$-chain that is composed of the corresponding sum of the restrictions of the map $\sigma_k$ to each of the faces $\phi_{i0}, \phi_{i1}$ of $I^k$:

$$\partial \sigma_k = \sum_i [\sigma_k(\phi_{i1}) - \sigma_k(\phi_{i0})] . \tag{2.4}$$

One could say that we require the boundary of $\sigma_k$ to be the image of the boundary of $I^k$.

One then extends the boundary operator $\partial$ to all of $C_*(M; R)$ by demanding that it be linear and that it agree with the boundary of any simplex in the formal sum. By convention, the boundary of $I^0 = \{0\}$ is zero, as well as the boundary of any 0-simplex and thus, any 0-chain. We put all of this together into a linear operator on the $\mathbb{Z}$-graded $R$-module $C_*(M; R)$:

$$\partial : C_*(M; R) \rightarrow C_*(M; R), \tag{2.5}$$

that we call the *boundary operator*. Since it takes $k$-chains to $(k-1)$-chains, one says that it has *degree* $-1$ relative to the $\mathbb{Z}$-grading. It is a basic property of the boundary operator that its square is zero:

$$\partial^2 = 0, \tag{2.6}$$

i.e., the boundary of a boundary is always zero.

As a simple example of the foregoing, the boundary of a 1-simplex $\sigma_1 : I \rightarrow M$ – i.e., a curve segment – is the oriented formal sum $\sigma_1(1) - \sigma_1(0)$ of its endpoints.

Although it might seems that the simplexes in $M$ define basic building blocks for the topology of $M$, at least to some extent, actually all of the topological information is contained in the particular definition of $\partial$ that pertains to $M$. It is an algebraic generalization of the construction of compact surfaces by identifying boundary points. In Fig. 1, we illustrate two simple examples of how the boundary operator defines the way that the simplexes are connected into chains in $M$:

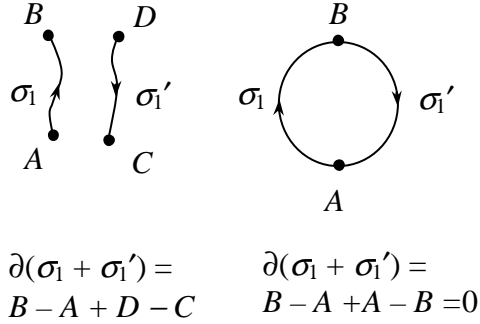

$$\partial(\sigma_1 + \sigma_1') = \qquad\qquad \partial(\sigma_1 + \sigma_1') =$$
$$B - A + D - C \qquad\qquad B - A + A - B = 0$$

Fig. 1. How the boundary operator defines the
way that simplexes are connected into chains.

When the boundary of a $k$-chain is zero one calls it a *k-cycle*, and we denote the $R$-module of all $k$-cycles by $Z_k(M; R)$. For instance, if one decomposes a loop in $M$ into a finite sum of curve segments then one can easily see how it has a vanishing boundary, since all pairs of endpoints appear twice, and with opposite signs.

Similarly, when a $k$-chain is the boundary of some $(k+1)$-chain one calls it a *k-boundary* and denotes the $R$-module of all $k$-boundaries by $B_k(M; R)$. Clearly, from (2.6), any $k$-boundary is a $k$-cycle; i.e., $B_k(M; R)$ is a sub-module of $Z_k(M; R)$.

The topology of $M$ enters the discussion when one asks whether all $k$-cycles in $M$ are $k$-boundaries, which bears heavily upon the idiosyncrasies of the operator $\partial$ for $M$ in particular. The quotient module $H_k(M; R) = Z_k(M; R) / B_k(M; R)$ is defined to the *homology R-module* of $M$ in dimension $k$. One can either think of its elements as the translates of $B_k(M; R)$ by all non-bounding $k$-cycles, or equivalence classes of $k$-cycles under the equivalence relation of *homology*. In particular, two $k$-cycles $c_k$ and $c_k'$ are called *homologous* iff they differ by the boundary of a $(k+1)$-chain:

$$c_k \sim c_k' \text{ iff} \qquad c_k - c_k' = \partial c_{k+1} \qquad\qquad \text{for some } c_{k+1} \in C_{k+1}(M : R). \qquad (2.7)$$

Whereas the $R$-modules $C_k(M; R)$, $Z_k(M; R)$, and $B_k(M; R)$ will generally be intractably high-dimensional, the homology module $H_k(M; R)$ has a generator for each "$k$-dimensional hole" in $M$, loosely speaking. When the coefficient ring is $\mathsf{R}$ (more generally: a principal ideal domain) there are no torsion factors in $H_k(M; R)$ and it is a vector space with a basis element for each of the aforementioned $k$-holes. For example, the punctured plane $\mathsf{R}^2 - \{0\}$ has any loop that encircles the origin – or rather, its homology class – as the generator of $H_1(M; R)$, which is then one-dimensional.

Since modern physics occasionally considers the way that topologically inequivalent manifolds can sometimes be interpolated by manifolds of one higher dimension – i.e., cobordisms – it is illuminating to show that the constructions of cobordism merely generalize the constructions of homology to the case where the objects that are being interpolated are not necessarily all contained in the same space. Hence, in Fig. 2, we illustrate the way that the widely discussed "trouser manifold" that connects a single loop to two disjoint loops by way of its boundary components can just as well represent a 2-chain whose boundary is the difference between two 1-cycles. In other words, two $k$-cycles are homologous if there is a $k+1$-chain that interpolates them as boundary components in essentially the same way that cobordism tries to do with closed manifolds, instead of $k$-cycles. We leave it to the reader to elaborate on the details of labeling the four vertices, eight edges, and four faces, and defining the boundary operator – i.e., defining the identifications of edges and vertices – that make this rigorous.

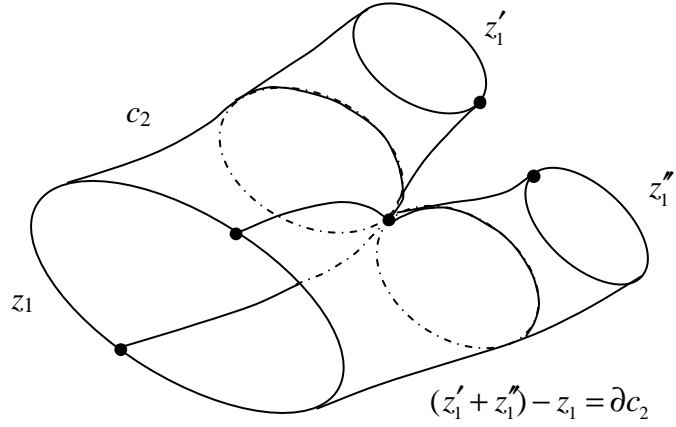

Fig. 2. The representation of the trouser manifold as a homology
(i.e., a 2-chain $c_2$) between the 1-cycles $z_1$ and $z_1' + z_1''$.

Since we shall be making use of homotopies of chains, we make a few remarks apropos of that objective. It is straightforward to see that when two $k$-chains are homotopic they will also be homologous. We note that the homotopy of chains resolves to a formal linear combination of homotopies of simplexes and examine the way that a homotopy of a two $k$-cycles produces a homology, namely, a $k+1$-chain whose boundary is the difference of the two $k$-cycles. One simply notes that if $\sigma_k$, $\sigma_k'$: $I^k \rightarrow M$ are two $k$-simplexes then a homotopy from one to the other is a continuous map $\sigma_{k+1}$: $I \times I^k \rightarrow M$ such that $\sigma_{k+1}(0, x) = \sigma_k(x)$ and $\sigma_{k+1}(1, x) = \sigma_k'(x)$. Hence, $\sigma_{k+1}$ is a $k+1$-simplex whose boundary contains $\sigma_k' - \sigma_k$, plus all of the lateral faces. What you have to convince yourself of (or simply look it up) is that when you put all of the $k$-simplexes together into two $k$-cycles the lateral faces of the connecting $k+1$-simplexes cancel out.

The converse statement that homologous $k$-cycles must be homotopic is not necessarily true, since homology is essentially a "coarser" level of equivalence than homotopy. In Fig. 3, we illustrate the manner by which a 1-cycle in the doubly punctured plane – namely, the sum $z_1 + z_1'$ of two circles that enclose one point, but not the other – can be homologous to the 1-cycle $z_1''$ without being homotopic. The reason that this does not contradict the Hurewicz isomorphism theorem is that the fundamental group $\pi_1(\mathsf{R}^2 - \{a, b\})$ of the doubly punctured plane is the free group on two generators [4], which is non-Abelian. One must then use the Abelianization of $\pi_1(\mathsf{R}^2 - \{a, b\})$, which is then a free Abelian group on two generators to obtain the isomorphism with $H_1(\mathsf{R}^2 - \{a, b\})$. As an example of a product in the free group that does not commute, consider the figure eight that is composed of two circles that intersect at a point. The order in which one describes one circle and then the other will determine two non-homotopic figure eights, but one can see that they are still homologous to the sum $z_1 + z_1'$.

_______________________
[4] For a discussion of these matters, see Massey [**5**].

In general, $R^2$ minus a finite set of points will have a fundamental group that is a free group with one generator for each point (i.e., some loop around it and none of the others) and the corresponding homology module in dimension one will be free Abelian with one generator for each point, as well. When one goes to $R^n$ with $n > 2$ the difference is that removing a point creates a "hole" in dimension $n - 1$.

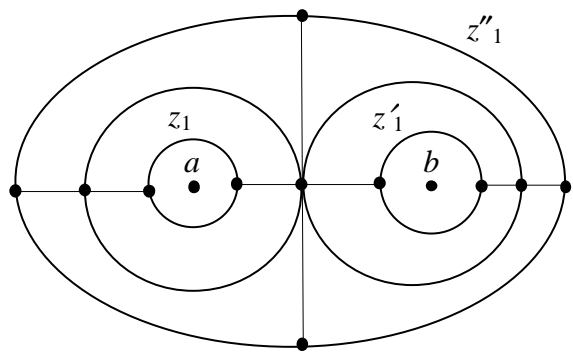

Fig. 3. How non-homotopic cycles can still be homologous.

The particular case of a homotopy of chains that we shall be most concerned with is that of a *deformation retraction*. If $A$ is a subspace of a topological space $X$ and $i: A \to X$ is the inclusion map then a deformation retraction of $X$ to $A$ is a map $r: X \to A$ such that the composition $i \cdot r: X \to X$ is homotopic to the identity on $X$. The subspace $A$ is then called a *deformation retract* of $X$. If one prefers to think of homotopies as things that take maps to maps then one must regard a deformation retraction as a homotopy from the identity map of $X$ to the inclusion of $A$ in $X$; this is more useful when the spaces $X$ and $A$ are defined by the carriers of two simplexes, for instance.

Although a deformation retraction is weaker than a homotopy equivalence, for which the other map $r \cdot i : A \to A$ would also have to be homotopic to the identity on $A$, nevertheless, the induced map in homotopy – hence, in homology – is an isomorphism in either case.

If $A$ consists of a single point then the deformation retraction is a *contraction* of $X$ onto $A$. In such a case the homotopy groups are all trivial, along with the homology groups, except for dimension 0, which is cyclic with one generator.

A simple example of a deformation retraction that we shall use is the retraction of $R^3$ onto the $z$-axis, which amounts to first factoring $R^3$ into $R^2 \times R^1$, and then contracting the $R^2$ factor to the origin by the radial contraction that was described above. Similarly, one can retract $R^3$ onto the $xy$-plane by contracting the $z$-axis to the origin. Although the homology is trivial in any of these cases, we shall generally be more concerned with the retraction of subspaces, anyway. For instance, one commonly retracts extended charge, mass, or current distributions on balls, tubes, and slabs or annuli, to distributions on points, curves, and surfaces, respectively. Hence, we shall use deformation retraction as a tool for eliminating the "homologically superfluous" dimensions in our field models.

## 2.2 Singular and de Rham cohomology

Since $\mathsf{R}$ is a field, we can simply characterize $H^k(M; \mathsf{R})$ as the dual space to the vector space $H_k(M; \mathsf{R})$; for more general rings, such as $\mathsf{Z}$, one must take the direct sum of this $\mathsf{R}$-module with a torsion summand.

One can first define a differentiable singular cubic $k$-*cochain* to be a linear functional on $k$-chains. However, one must keep in mind that since the cardinality of any basis for $C_k(M; \mathsf{R})$ equals the cardinality of the set of $k$-simplexes – i.e., it is potentially uncountably infinite – defining an isomorphism of $C_k(M; \mathsf{R})$ and $C^k(M; \mathsf{R}) = C_k(M; \mathsf{R})^*$ will not always be canonical. However, one can, at least define a useful linear injection by means of the map that takes any $k$-simplex $\sigma_k$ to its corresponding *characteristic functional* $\sigma^k$, which is defined by:

$$\sigma^k(\sigma'_k) = \begin{cases} 1 & \text{if } \sigma'_k = \sigma_k \\ 0 & \text{otherwise.} \end{cases} \tag{2.8}$$

Hence, any $k$-chain $c_k = \sum a_i \sigma_{k,i}$ corresponds to the $k$-cochain $c^k = \sum a_i \sigma^k_{,i}$.

We now introduce a common notation from homology theory, namely the bracket that pairs a linear functional, such as $c^k$, and a vector, such as $c^k$, with the scalar that equals the value of the functional when applied to the vector:

$$<.,.>: C^k(M; \mathsf{R}) \times C_k(M; \mathsf{R}) \to \mathsf{R}, \qquad (c^k, c_k) \mapsto <c^k, c_k> = c^k(c_k). \tag{2.9}$$

One can define the *coboundary* operator $\delta$ on cochains by a generalized "Stokes theorem" construction:

$$<\delta\alpha^k, c_{k+1}> = <\alpha^k, \partial c_{k+1}>, \qquad \text{for all } c_{k+1} \in C_{k+1}(M; \mathsf{R}). \tag{2.10}$$

Given this coboundary operator, one immediately defines cocycles, coboundaries, cohomologous cocycles, and the cohomology vector spaces $H^k(M; \mathsf{R}) = Z^k(M; \mathsf{R}) / B^k(M; \mathsf{R})$, in the obvious way.

In order to make the transition to de Rham cohomology direct, note that since the space $M$ that we are dealing with is a differentiable manifold, on which we can integrate $k$-forms over $k$-chains, we can associate a linear functional on $C_k(M; \mathsf{R})$ with every $k$-form $\alpha$, namely:

$$<\alpha, c_k> = \int_{c_k} \alpha. \tag{2.11}$$

Not every linear functional on $C_k(M; \mathsf{R})$ can be so represented, and, in particular, the evaluation functionals cannot. Of course, the non-existent kernel of the evaluation functional is what physics and mathematical analysis has always defined to be the *Dirac delta function*, a fact that will clearly be useful to recall when we discuss charge and charge densities.

If $\Lambda^k(M)$ represents the vector space of exterior differential $k$-forms on $M$ then we call the vector space $Z^k(M)$ of closed $k$-forms the space of *de Rham k-cocycles* and the vector space $Z^k(M)$ of exact $k$-forms the space of *de Rham k-coboundaries*. The quotient vector space $H^*_{dR}(M) = Z^k(M) / B^k(M)$ is then called the *de Rham cohomology* vector space in dimension $k$. Its elements are equivalence classes of closed form that differ by an exact form. The gist of de Rham's theorem, at least in the present context of differential singular cubic homology, is then that $H^*_{dR}(M)$ is isomorphic to $H^k(M; \mathsf{R})$ – i.e., the differentiable singular cubic cohomology with real coefficients − for each dimension $k$.

The main steps in the proof come from Stokes's theorem, when given the form:

$$< \alpha, \partial c_{k+1} > = < d\alpha, c_{k+1} > , \qquad (2.12)$$

and the Poincaré lemma. This shows that the operators $d$ and $\partial$ are, in a sense, adjoint to each other with respect to the given bilinear pairing.

For the sake of completeness, we point out that the $k$-chain $c_{k+1}$ also defines a linear functional on the vector space $\Lambda^k(M)$ of $k$-forms on $M$ by way of the pairing (2.11). These linear functionals on $\Lambda^k(M)$ are what de Rham [6] referred to as *currents*, although we shall not use this approach in the present study. In effect, they represent elements of $C_k(M; \mathsf{R})$ by elements of the dual space to $\Lambda^k(M)$, which is itself the dual space to $\Lambda_k(M)$. In a later section, we shall see that on an orientable manifold one can define a "de Rham homology" over $\Lambda_*(M)$ more directly than by means of elements of the dual of a dual space.

## 2.3 Poincaré duality

The introduction of a volume element on $M$ into the field theory is very immediately related to the homology of the source complex since one cannot define a divergence operator without one and the divergence plays a key role in the context of source currents. Since it is not universally acknowledged that one can define a divergence operator for $k$-vector fields on an orientable manifold without needing to introduce a metric or that the resulting operator allows one to define a real homology that is essentially dual to the de Rham cohomology – minus the ring structure – we shall briefly summarize those constructions.

Let $M$ be an orientable $n$-dimensional differentiable manifold and let $\varepsilon \in \Lambda^n(M)$ be a unit-volume element on $T^*(M)$; i.e., a non-zero $n$-form. In a local coordinate chart $(U, x^\mu)$ the component expression for $\varepsilon$ takes the form:

$$\varepsilon = dx^1 \wedge \dots \wedge dx^n = \frac{1}{n!} \varepsilon_{i \dots j} \, dx^i \wedge \dots \wedge dx^j , \qquad (2.13)$$

in which $\varepsilon_{i \dots j}$ represents the completely anti-symmetric Levi-Civita symbol with $n$ indices.

This unit-volume element on $T^*(M)$ is associated with a corresponding unit-volume element $\mathbf{\varepsilon} \in \Lambda_n(M)$ on $T(M)$, which is then a non-zero $n$-vector field on $M$ that relates to $\varepsilon$ by way of the defining property:

$$\varepsilon(\boldsymbol{\varepsilon}) = 1. \tag{2.14}$$

The local component form of $\boldsymbol{\varepsilon}$ is then:

$$\boldsymbol{\varepsilon} = \partial_1 \wedge \dots \partial_n = \frac{1}{n!} \varepsilon^{i\dots j} \partial_i \wedge \dots \partial_j. \tag{2.15}$$

An immediate consequence of the existence of $\varepsilon$ and $\boldsymbol{\varepsilon}$ is the existence of the isomorphisms of *Poincaré duality:*

$$\# : \Lambda_k(M) \to \Lambda^{n-k}(M), \qquad \mathbf{a} \mapsto i_{\mathbf{a}}\varepsilon, \tag{2.16a}$$

$$\#^{-1} : \Lambda^k(M) \to \Lambda_{n-k}(M), \qquad \alpha \mapsto i_{\boldsymbol{\varepsilon}}\alpha. \tag{2.16b}$$

(In these expressions, we intend that the combination of multivectors and differential forms is by the interior product.) The local expressions for the components are:

$$(\#\mathbf{a})_{\mu\dots\nu} = \frac{1}{k!} a^{\kappa\dots\lambda} \varepsilon_{\kappa\dots\lambda\mu\dots\nu}, \tag{2.17a}$$

$$(\#^{-1}\alpha)_{\mu\dots\nu} = \frac{1}{k!} \alpha_{\kappa\dots\lambda} \varepsilon^{\kappa\dots\lambda\mu\dots\nu}. \tag{2.17b}$$

These isomorphisms are actually of projective geometric origin since one can use decomposable $k$-vector fields and decomposable $k$-forms to represent $k$-dimensional vector subspaces of $\mathsf{R}^n$, either directly or by way of their annihilating subspaces. This sort of duality says essentially that the same $k$-plane in $\mathsf{R}^n$ can either be represented directly by a decomposable $k$-vector or indirectly by the annihilating subspace of an $n-k$-form.

It is important to point out that although the $\#$ map is a linear isomorphism of vector spaces, it is not actually an isomorphism of exterior algebras since $\#(\mathbf{a} \wedge \mathbf{b})$ is an $(n-k-l)$-form, whereas $\#\mathbf{a} \wedge \#\mathbf{b}$ is a $(2n-k-l)$-form.

## 2.4 De Rham homology

When one has an orientable, oriented manifold $M$ with a chosen unit-volume element $\varepsilon$ one can use Poincaré duality and the exterior derivative operator on differential forms to define a divergence operator on multivector fields that generalizes both the divergence operator on vector fields that one first encounters in vector calculus, as well as the codifferential operator on differential forms that one defines by means of the Hodge duality operator. In particular, the divergence operator $\delta: \Lambda_*(M) \to \Lambda_{*-1}(M)$ is defined by:

$$\delta = \#^{-1}d\#. \tag{2.18}$$

It is a linear operator with the property:

$$\delta^2 = 0 \, , \tag{2.19}$$

and its local expression in terms of the components a vector field $\mathbf{v} = v^\mu \, \partial_\mu$ is the classical one:

$$\delta \mathbf{v} = \frac{\partial v^\mu}{\partial x^\mu} \, . \tag{2.20}$$

One defines a $k$-vector $\mathbf{a}$ to be a *de Rham k-cycle* if it has zero divergence and a *de Rham k-boundary* if it is the divergence of some $(k + 1)$-vector. If we denote the vector space of de Rham $k$-cycles by $Z_k(M)$ and the vector space of de Rham $k$-boundaries by $B_k(M)$ then we can define the quotient vector space $H_*^{dR}(M) = Z_k(M) \, / \, B_k(M)$ to be the *de Rham homology* vector space of $M$ in dimension $k$. It follows from Poincaré duality and de Rham's theorem that this vector space is isomorphic to the $k^{th}$ differentiable singular cubic homology module when one uses the real numbers for one's coefficient ring. As an aid to one's intuition, one might consider that if one regards a differentiable singular cubic $k$-chain $\sigma_k$ as a restriction of a $k$-dimensional submanifold in $M$ then the $k$-dimensional tangent space to $\sigma_k$ can be represented by an equivalence class of decomposable $k$-vector fields with compact support under multiplication by a scalar function on $M$.

Unlike the exterior derivative operator, the divergence operator is not an anti-derivation; i.e., one does not have $\delta(\mathbf{a} \wedge \mathbf{b}) = \delta\mathbf{a} \wedge \mathbf{b} + (-1)^p \, \mathbf{a} \wedge \delta\mathbf{b}$, since # is not an isomorphism of exterior algebras. Hence, the exterior product does not "descend to homology" to define a ring structure on $H_*^{dR}(M)$, as one obtains for de Rham cohomology.

However, one should keep in mind that the vector space $\mathsf{X}(M)$ of vector fields on $M$ has a Lie algebra defined over it, and the divergence operator does have the property that the Lie bracket of two 1-cycles is a 1-cycle; i.e., if $\delta\mathbf{v} = 0$ and $\delta\mathbf{w} = 0$ then $\delta[\mathbf{v}, \mathbf{w}] = 0$. Hence, the 1-cycles − or *conserved currents* − on $M$ form a Lie subalgebra of $\mathsf{X}(M)$. Indeed, such vector fields are simply the infinitesimal generators of local volume-preserving diffeomorphisms of $M$, and the global volume-preserving diffeomorphisms form a Lie group. Although the Lie algebra on $\mathsf{X}(M)$ can be extended to a non-trivial Lie algebra on $\Lambda_*(M)$ [5], whether this Lie algebra descends to homology to define a Lie algebra on $H_*(M; \mathrm{R})$ is not entirely clear to the author .

## 3 Topological field theories

### 3.1 Charge, flux, and current

Customarily, one associates charge with a region of space either by way of a charge density function that is defined on the region or a total charge that is contained in the

---

region. If we restrict ourselves to regions of space that can be described by the carrier of a differentiable singular $k$-chain $c_k$ then if the association of total charge $Q$ with chains is assumed to be linear, we can characterize $Q$ as a differentiable singular cubic $k$-cochain, and write its value when applied to a $k$-chain $c_k$ as either $Q[c_k]$ or $<Q, c_k>$.

In order to resolve this linear functional on $k$-chains into an integral functional, one must define a $k$-form $\rho$ that serves as the *charge density $k$-form* for the functional $Q$ because, by definition:

$$Q[c_k] = \int_{c_k} \rho \, . \tag{3.1}$$

When the $k$-form $\rho$ is closed, the cochain $Q[]$ will be vanish on any $k$-boundary, which will make it a cocycle. Furthermore, if $\rho$ is exact then $Q[]$ will vanish on any $k$-cycle, which will make it a coboundary.

One can use the total charge functional to define a total charge $k$-chain $Q_k$ that corresponds to any $c_k = \sum a_i \, \sigma_{k,i}$ by way of:

$$Q_k = \sum a_i \, Q[\sigma_{k,i}] \, \sigma_{k,i} \, . \tag{3.2}$$

In other words, we are multiplying each component $k$-simplex by the total charge that is distributed over it.

One can even apply this construction in the case where the total charge functional does not have a differential form as a kernel, such as a linear combination of characteristic functions for a finite set of $k$-simplexes. In particular, a set of $N$ point charges $Q_i$, $i = 1, \ldots, N$ whose points are described by $N$ corresponding 0-simplexes $\sigma_{0,i}$, $i = 1, \ldots, N$ can be associated with either the total charge 0-chain $\sum Q_i \sigma_{0,i}$, or the total charge 0-cochain $\sum Q_i \sigma^0{}_{,i}$, in which $\sigma^0{}_{,i}(\sigma_{0,j}) = \delta_{ij}$.

Suppose $M$ is orientable, oriented, and given a choice of unit volume element $\varepsilon$. If we agree to call a vector field $\mathbf{v}$ with vanishing divergence a *conserved current* then it is natural to question the physical meaning of its Poincaré dual #$\mathbf{v}$. Since it is an $n-1$-form, it can be integrated over a compact orientable $n-1$-dimensional submanifold of $M$; more to the point, it can be integrated over a differentiable singular cubic $n-1$-chain $c_{n-1}$. In physics, it is customary to identify the resulting number:

$$\Phi_\mathbf{v}[c_{n-1}] = <\#\mathbf{v}, c_{n-1}> = \int_{c_{n-1}} \# \, \mathbf{v} \tag{3.3}$$

with the *total flux* of $\mathbf{v}$ through $c_{n-1}$. From the previous considerations, is natural to identify the $n-1$-form #$\mathbf{v}$ with the *flux density* of the vector field $\mathbf{v}$. Hence, the linear functional $\Phi_\mathbf{v}[]$ is an $n-1$-cochain. Furthermore, if we assume that $\mathbf{v}$ has zero divergence, then #$\mathbf{v}$ is closed, and one has that the resulting linear functional $\Phi_\mathbf{v}[]$ is a cocycle, since it clearly vanishes when $c_{n-1}$ is a boundary. When $\mathbf{v}$ is the divergence of a bivector, $\Phi_\mathbf{v}[]$ will vanish whenever $c_{n-1}$ is a cycle, and $\Phi_\mathbf{v}[]$ will be a coboundary.

One generally tends to think of a current, such as electric current, as something that is represented by a divergenceless vector field $\mathbf{J}$ of the form $\rho\mathbf{v}$ on a spatial region $\mathsf{S}$ of compact support, where $\rho$ is a density function (e.g., mass, charge, etc.) on $\mathsf{S}$ and $\mathbf{v}$ is a velocity vector field that describes the motion of the quantity whose density is described by $\rho$. Furthermore, although there is nothing to say that the current is not flowing in an inhomogeneous region where the conductance varies from point to point in such a way that $\mathbf{J}$ is not spatially constant, nevertheless, the case of a current whose magnitude does not vary spatially is also very physically meaningful, if only as the ideal case of a perfect conductor.

Furthermore, the support $\mathsf{S}$ of a current might takes the form of a network – often, just a loop – with a finite number of branches $b_i$, $i = 1, …, N$ and vertices $v_j$, $j = 1, …., M$, and an incidence law that associates them; i.e., a 1-dimensional simplicial complex [6].

If the current $I_i$ in the branch $b_i$ is constant along the branch then one can give physical meaning to the formal product $I_i b_i$. Since stable currents flow only in loops, one can also specify that this complex is closed. Hence, the set $\{z_l, l = 1, …, L\}$ of all elementary loops (i.e., not the sum of simpler loops) that can be formed from the branches of the network define a set of generators for $Z_1(\mathsf{S}; \mathsf{R})$. Moreover, since one generally presumes that none of the loops bound any two-dimensional region of space that can still be described as belonging to the source – i.e., = 0 – one then also has that the set $\{z_l, l = 1, …, L\}$ represents a set of generators for $H_1(\mathsf{S}; \mathsf{R})$. One must notice that, from a physical standpoint, the assumption that the 2-chains of $\mathsf{G}$ that are bounded by loops in $\mathsf{S}$ do not contribute to the source current breaks down when there are time-varying magnetic fields present.

Physically, charge and current are related concepts, since current is a time rate of flow of charge. One associates a charge $Q_j(t)$ – which we regard as a differentiable time-varying function – with each vertex $v_j$ by defining the total charge 0-cocycle $Q_j v^j$, where $v^j$ is the $j^{\text{th}}$ member of the reciprocal basis for $C^0(\mathsf{S}; \mathsf{R})$ that is defined by $v^j(v_j) = \delta^i_j$. One associates a current $I_i$ – which need only be a constant real number, for the present purposes – with each branch $b_i$ by forming the total current 1-cochain $I_i b^i$, where $b^i$ is the reciprocal basis to $b_i$, along with the total current 1-chain:

$$I_{\text{tot}} = \sum_j I_j b_j \ . \tag{3.4}$$

One notes that the (singular) coboundary $\delta v^j$ of the vertex $v^j$ is:

$$\delta v^j = \sum_j a^i_j b^j \ , \tag{3.5}$$

where the "incidence matrix" $a^i_j$ is defined by:

$$a^i_j = v^i(\partial b_j) = \begin{cases} \pm 1 & \text{if } \partial b_j = \pm v^i \\ 0 & \text{otherwise} \end{cases} \ . \tag{3.6}$$

With these notations, we see that Kirchhoff's law of currents, which is equivalent to the conservation of charge, takes the purely cohomological form:

$$\frac{dQ_i}{dt} = <\delta v^i, I_{tot}> = <v^i, \partial I_{tot}>,\tag{3.7}$$

which gives the signed sum of the currents from the branches that have $v^i$ as one endpoint, and when the charge at any vertex is constant in time, the statement that the sum of the currents at that vertex is zero takes the form:

$$0 = <\delta v^i, I_{tot}> = <v^i, \partial I_{tot}> .\tag{3.8}$$

Hence, since this is true for all $i$, the 1-chain $I_{tot}$ is a 1-cycle.

We note, in passing, that this methodology can be applied directly to a system of masses coupled by forces. If linear momentum plays the role of charge and force the role of current then the analogue of Kirchhoff's law of current (3.7) is Newton's second law of motion, while (3.8) describes the equilibrium state of the system.

For the sake of completeness, we point out that if one associates a differentiable time-varying potential function $V(t)$ to each vertex and a potential difference $\Delta V_j$ to each branch in a similar way then Kirchhoff's law of voltages, which is equivalent to conservation of energy, takes the form:

$$\frac{dV_k}{dt} = <\Delta V_{tot}, \partial c_k> = <\delta \Delta V_{tot}, c_k>,\tag{3.9}$$

in which – by abuse of notation − $c_k$ is a 1-chain. When the total energy is conserved, one sees that the sum of the potential differences around any loop is zero; i.e., $\Delta V_{tot}$ is a 1-cocycle.

## 3.2 Charge and flux quantization

So far, we have been concerned only with homology and cohomology with coefficients in a field, namely $\mathbb{R}$. Since the electrical charge that is found in Nature is known to be reducible to a (generally very large) sum of elementary charges of magnitude $\pm e$ (unless one includes quarks), one might prefer to regard any charge $Q$ as being of the form $Ze$, where $Z \in \mathbb{Z}$. As we have been using $Q$ as a coefficient in various chains or cochains, this suggests that at a fundamental level we should properly be using coefficients in the ring $\mathbb{Z}$. This has advantages and disadvantages, since homology with integer coefficients is richer in structure, because it can include torsion factors that disappear in real homology, but such richness cannot usually be represented in terms of de Rham cycles or cycles.

The fundamental question to both physics and topology is: "In what sort of field theory is a reduction from $\mathbb{R}$ to $\mathbb{Z}$ canonical, in some sense?" One finds that generally the integrality of charges originates in homotopy considerations, which then imply corresponding consequences in homology or cohomology.

Whenever one is evaluating a $k$-cocycle over a $k$-cycle that represents a simplicial decomposition of a (diffeomorphic image of a) $k$-sphere in a manifold $M$, one must always keep in mind that there might very well be more than one homotopy class of $k$-spheres in $M$; i.e., one might have $\pi_k(M) \neq 0$. In particular, one might have $\pi_k(M) \cong \mathsf{Z}$, where one might think of the integer that gets associated with each homotopy class as the winding number of the map that takes $S^k$ to its image in $M$.

In such a situation – for instance, when $M = \mathsf{R}^3 - \{0\}$, which has the same homotopy type as $S^2$ – the charge $Q = <\rho, z_k>$ that is associated with the generator $[z_k]$ of $\pi_k(M)$, namely, the homotopy class that is associated with 1, takes on the character of a fundamental charge, such that all of the other charges will be integer multiples of $Q$. Since $\pi_k(S^k) \cong \mathsf{Z}$ for any $k > 0$, this is the most common way of trying to introduce integrality into modern field theories, such as gauge field theories.

One can even think in terms of quantized – i.e., discrete – currents in some cases, such as the current in a conducting loop that consists of $N$ turns of a wire that carries a current $I$; i.e., a $\mathsf{Z}$-cycle $Nz_1$ would represent $N$ copies of a current loop $z_1$ and the corresponding current cocycle would take the form $NI$.

### 3.3 Topological field theories

Ultimately, the objective of any field theory should be to account for the coupling of the information in a field source to the information in the field itself. When one examines the motivating examples for the sort of field that we shall a "topological field theory", such as electrostatics, one finds that the field equation for the field does not actually involve the source distribution explicitly, and that the role of the charge that is associated with the field source is introduced only by way of the association of the total flux through a surface with the total charge "contained" in the surface. Although this usually seems to follow from Gauss's law, one finds that on closer inspection of the Coulomb case, it does not follow in this way because, for one thing, Gauss's law is *inapplicable* when the surface does not bound, and for another, direct calculation shows that the divergence of the field is zero. Hence, the step at which one sets the volume integral of the divergence equal to the total charge is incorrect. Since this is equivalent to the equation $\delta \mathbf{E} = \rho$, one sees that this equation is itself misleading, even when one sets $\rho = Q\delta^0$. Hence, our objective is to reconstruct the overall effect of the misapplication of Gauss's law while paying careful attention to the topological details.

For the time being, we simply assume that we have an $n$-dimensional space or spacetime manifold $M$, a closed subset $\mathsf{S}$ of $M$ that we call the set of *source points*, and its open complement $\mathsf{G}$, which we call the set of *field points*. This implies that $\mathsf{G}$ will be a submanifold of $M$ of dimension $n$. Furthermore, we assume that $\mathsf{G}$ is dense in $M$: i.e., $M = \bar{\mathsf{G}}$. This, and the fact that $\mathsf{S} = M - \mathsf{G}$, imply that $\mathsf{S}$ represents all of the non-generic limit points of the set $\mathsf{G}$. Hence, every neighborhood of a source point will contain a field point.

Along with the decomposition of $M$ into $\mathsf{G}$ and $\mathsf{S}$, we associate two spaces [7]: a *field space* $\Gamma(\mathsf{G})$ and a space of *source distributions* $\$(\mathsf{S})$. What we will be defining as *topological field theories* are field theories in which:

*a*) The closed set $\mathsf{S}$ is the carrier of a differentiable singular cubic $k$-cycle $z_k$.

*b*) The space of source distributions is $H^k(\mathsf{S}; \mathsf{R})$.

*c*) The field space is $H^m(\mathsf{G}, \mathsf{R})$ for some $m \geq k$.

*d*) The coupling of one to the other takes the form of a linear isomorphism:

$$H^k(\mathsf{S}; \mathsf{R}) \rightarrow H^m(\mathsf{G}, \mathsf{R}). \tag{3.10}$$

For the sake of physics, restricting oneself to source complexes that can be expressed as homogeneous closed chains is not a severe restriction, since it at least includes the common elementary examples. For instance, $\mathsf{S}$ may be as elementary as a finite set of isolated points, which represents a zero-dimensional cycle, or perhaps a finite set of non-intersecting closed curves, which then represents a one-dimensional cycle. In the latter case, we are including the possibility that the curves close "at infinity," so we are assuming that the space in which the fields are defined is either compact or compactified by a point at infinity or a hyperplane at infinity. This is simply a different form of the usual physical requirement that a physical field must vanish at infinity. Furthermore, one might have two-dimensional source complexes, which can represent surface charge distributions or wave discontinuity surfaces, and even three-dimensional source complexes.

However, since finite chains can only describe compact subspaces of $M$, one must either confine all of ones attention to compact restrictions of non-compact global constructions – such as infinitely-extended lines and planes – or use only compactified $M$'s. In the case of line and plane sources, one must remember that regarding them as infinitely extended was a simplifying assumption to begin with since such things as real world wires and capacitor plates are always finite in their extent, so regarding them as infinite is just a way of neglecting the inhomogeneities at their boundaries. We shall see in a later section that generally the non-compact sources can be treated as homologically equivalent to corresponding compact sources.

In any event, by definition, a closed source complex $\mathsf{S}$ defines a homology class $[\mathsf{S}] \in H_k(\mathsf{S}; \mathsf{R})$. Hence, the space of source distributions consists of all possible total charge functionals that can be applied to the points of $\mathsf{S}$; i.e., to the $k$-simplexes that triangulate it by way of $z_k$. When $\mathsf{S}$ is more than zero-dimensional, these functionals can be defined by the restrictions of corresponding closed $k$-forms to $\mathsf{S}$, which then represent possible charge densities.

It might seem that a topological field theory is devoid of field equations in the usual sense of systems of partial differential equations, since we seem to be more concerned with algebraic relationships between source cycles or cocycles and field cocycles. However, one must remember that in order to find a field that represents a de Rham $m$-cycle, one must find an $m$-vector field whose divergence vanishes and to find a field that represents a de Rham $k$-cocycle, one must find a $k$-form whose exterior derivative vanishes. Hence, the differential equations of our field theory might generally take either the form:

$$\delta \mathbf{J} = 0 \tag{3.11}$$

or the form:

$$d\phi = 0. \tag{3.12}$$

---

[7] At this point, we are being casual and heuristic in our usage of the word "space" since what we are thinking in terms of is a space of sections of some appropriate fiber bundle, which might not be a vector bundle, as in the case of some nonlinear field theories, but we will not use the spaces $\$(\mathsf{S})$ and $\Gamma(\mathsf{G})$ any further in the present study, anyway.

### 3.4 Field/source duality

In this section, we deal with the fourth axiom that we introduced above, which has the effect of coupling the total charge cocycle for a given source to associated total flux cocycle for the field that it generates. This axiom generalizes the usual association that is supposed to result from the misapplication of Gauss's law to cycles that do not bound. Hence, we will have to reconstruct its basic effect – of equating topological information in the field to topological information in the source – "by hand," so to speak.

One should observe that the association of cohomology classes with other such classes is quite broad in its scope as far as physical interpretations is concerned. In particular, one is not associating particular charge densities with particular fields, at all, but only large equivalence classes. For instance, as we shall discuss later in more detail, in the case of a single point source in $\mathsf{R}^3$ any total charge cocycle takes the form $Q[\delta_P]$, where $Q \in \mathsf{R}$ and $[\delta_P]$ is a generator for the one-dimensional vector space $H^0(\mathsf{S}; \mathsf{R})$, and the total flux cocycle takes the form $\Phi[\Sigma]$, where $\Phi \in \mathsf{R}$ and $[\Sigma]$ is a generator for the one-dimensional vector space $H^2(\mathsf{G}; \mathsf{R})$. The field/source duality can then be defined by the association of $[\delta_P]$ with $[\Sigma]$.

Of course, there would be something unsatisfying and unconvincing about our substitute for the usual association of total flux and total charge if it were merely an *ad hoc* prescription. Hence, we need to look into the possibility that the field/source duality isomorphism might have some deeper topological roots than a mere coincidence of generators.

First, we observe that something topologically interesting can happen when one puts fields and sources into complementary subspaces: things that might have been elementary when the two subspaces were together can become non-trivial when the pieces are separated. For instance, although $\mathsf{R}^2$ is a contractible topological space, hence, its homotopy and homology groups are trivial, the punctured plane has a non-vanishing fundamental group, as well as non-vanishing one-dimensional homology and cohomology, just as removing the origin from $\mathsf{R}$ gives it non-trivial homotopy and homology in dimension zero. In general, removing the origin from $\mathsf{R}^n$ ($n > 2$) produces a space that has $S^{n-1}$ as a deformation retract, which means that its first non-trivial homotopy group is in dimension $n-1$, and by the Hurewicz isomorphism theorem (see [1]), $\pi_{n-1}(S^{n-1})$ will be isomorphic to $H^{n-1}(S^{n-1})$. Moreover, they are both isomorphic to $\mathsf{Z}$, so they both have one generator. Note that since the point that we removed has non-vanishing cohomology only in dimension zero, which is also isomorphic to $\mathbb{Z}$, we could only expect to couple the generator of $H^0(\{0\}; \mathbb{Z})$ to $H^{n-1}(\mathbb{R}^n - \{0\}; \mathbb{Z})$.

At an elementary level, one can always define a $k$-cycle that does not bound a $k+1$-chain by starting with a $k+1$-simplex and removing the interior. Similarly, when one removes a $k+1$-simplex $\sigma_{k+1}$ from $\mathbb{R}^n$ one produces a topological space whose homology module in dimension $k$ has a generating homology class that is represented by any $k$-sphere in $\mathbb{R}^n$ that projects inward onto $\partial\sigma_{k+1}$. Hence, although the generators of $H_k(M)$ are often loosely referred to as "$k$-dimensional holes in M" they are really more like the boundaries of such things.

Furthermore, since any simplex is contractible to a point the space that is obtained by removing $\sigma_{k+1}$ from $\mathbb{R}^n$ is homologically equivalent (i.e., isomorphic) to the one that obtained by removing the origin from $\mathbb{R}^n$. Note that one could just as well have removed the interior of an $n$-simplex, since ultimately the effect on homology was to introduce a generator in dimension $n-1$. We shall see that this is related to the fact that the external field of a spherically symmetric charge distribution is the same as the field of a point charge.

In order relate the topological information in the source complex to the topological information in the field complex, what we need to find is a natural way of bridging the gap that we create by separating the source points from the field points. As a heuristic motivation, consider that in the example of the punctured plane the unit circle can be represented by a 1-cycle either in $H_1(\mathbb{R}^2, \mathbb{R})$ or in $H_1(G, \mathbb{R})$, but since $H_1(\mathbb{R}^2, \mathbb{R}) = 0$, it is a boundary in $\mathbb{R}^2$, namely, the boundary of the unit disc about the origin, though it is not a boundary in $G$, the punctured plane, since the region inside the unit sphere is no longer a 2-chain because it is not compact.

Suppose $G$ has non-vanishing homology in dimension $k$. We then define a $k$-cycle $z_k \in Z_k(G; \mathbb{R})$ to be a *partial $k$-boundary* if it is a boundary in $M$, but it is not a boundary in $G$. The partial boundaries define a vector subspace of $Z_k(G; \mathbb{R})$ that we denote by $Z_{k*}(G; \mathbb{R})$. Our ongoing example of the unit circle in the punctured plane would be an example of a partial 1-boundary.

Let $z_k$ be a partial $k$-boundary in $G$ that bounds a $k+1$-chain $c_{k+1} \in Z_{k+1}(M; \mathbb{R})$ that has an $m$-cycle $z_m \in Z_m(S; \mathbb{R})$ as a deformation retract. If the deformation retraction that takes $c_{k+1}$ to $z_m$ is denoted by $r$ then we can describe this scenario by the following set of maps:

$$z_k \xleftarrow{\ \partial\ } c_{k+1} \xrightarrow{\ r\ } z_m$$

For example, $z_k$ could be the unit sphere in $\mathbb{R}^3$, $c_{k+1}$, the closed unit ball that it bounds, and $z_m$, the origin. We shall call a $k+1$-chain such as $c_{k+1}$ a *connecting $k+1$-chain*.

When such a connecting chain exists, one has an association of partial $k$-boundaries in $Z_{k*}(G; \mathbb{R})$ with $m$-cycles in $Z_m(S; \mathbb{R})$. Since that induces a corresponding homomorphism of $H_m(S; \mathbb{R})$ into $H_k(G; \mathbb{R})$, the key issue to address next is when that homomorphism is an isomorphism. Clearly, this is the case only if every homology class in $H_k(G; \mathbb{R})$ contains a partial boundary. Moreover, there must be a connecting $k+1$-chain between every generator of $H_k(G; \mathbb{R})$ and some generator of $H_m(S; \mathbb{R})$ in an invertible manner. Since the general problem of finding connecting chains appears quite open-ended at this point, in this study, we shall content ourselves to describing them for specialized situations.

Given an isomorphism of $f$: $H_m(\mathsf{S}; \mathsf{R}) \to H_k(\mathsf{G}; \mathsf{R})$, in order to define an isomorphism of $H^m(\mathsf{S}; \mathsf{R})$ with $H^k(\mathsf{G}; \mathsf{R})$, one need only pull back the linear functionals on $H_k(\mathsf{G}; \mathsf{R})$ to linear functionals on $H^m(\mathsf{S}; \mathsf{R})$ by way of $f$; i.e., $f^*$: $H^k(\mathsf{G}; \mathsf{R}) \to H^m(\mathsf{S}; \mathsf{R})$ is the desired isomorphism.

If we wish to take a direct route from $H^k(\mathsf{G}; \mathsf{R})$ to $H^m(\mathsf{S}; \mathsf{R})$, we need to first define a *partial coboundary* to be a real $k$-cocycle $z^k$ in $M$ such that it is a coboundary in $H^k(M; \mathsf{R})$, but not in $H^k(\mathsf{G}; \mathsf{R})$. A connecting cochain for a partial boundary $z^k \in H^k(\mathsf{G}; \mathsf{R})$ and an $m$-cocycle $z^m \in H^m(\mathsf{S}; \mathsf{R})$ is a $k$–1-cochain $c^{k-1} \in H^{k-1}(M; \mathsf{R})$ such that $z^k = \delta c^{k-1}$ and there is a deformation retraction from $c^{k-1}$ to $z^m$.

### 3.5 Equivalent field/source configurations

Since the fundamental level at which we are coupling sources to fields is the level at which we are coupling the total flux cocycle that is associated with a field with the total charge cocycle that is associated with its source, it is reasonable to define two source/field configurations as being equivalent when the total flux cocycle for an analogous field is associated with the total charge cocycle for the analogous source. In other words, if $\phi$ is a closed flux density $k$-form on $\mathsf{G}$ whose total flux $k$-cocycle is $\Phi_\phi[.]$ and $\rho$ is a closed charge density $m$-form on $\mathsf{S}$ whose total charge $m$-cocycle is $Q_\rho[.]$ then under an association of $\phi$ with a closed $k$-form $\phi'$ on $\mathsf{G}'$ and $\rho$ with a closed $m$-form $\rho'$ on $\mathsf{S}'$, if $\Phi_\phi[.]$ is associated with $Q_\rho[.]$ to begin with, one must have that $\Phi_{\phi'}[.]$ is associated with $Q_{\rho'}[.]$ as a consequence.

Some equivalences are automatic, as consequences of the use of cohomology. For instance, two homeomorphic source cycles $z_m$ and $z'_m$ – i.e., their carriers – will also be homotopically and homologically equivalent. Hence, we are not really concerned with the locations of the source points in $M$ or the overall geometrical shape of the source cycle.

Furthermore, since field/source duality only associates cohomology classes, and not their representative cocycles, another automatic equivalence is that cohomologous field $k$-forms will define the same cohomology class, and similarly for cohomologous charge densities.

Physically, what we are introducing in the form of homologous source complexes includes the freedom to translate and deform the source complex in a fairly general way without altering anything fundamental in the definition of the field theory. However, one must keep in mind that, in reality, physical field theories eventually have to introduce more rigid sort of geometrical considerations, and one finds that the solution of the electrostatic potential for a spherical boundary is computationally simpler than the solution of the same problem for an ellipsoidal boundary, even if the two boundaries are homologous. However, looking at that same situation in the reverse order, one realizes that homology classes of field theories might sometimes contain "canonical" representatives that are more elementary, in some physically useful sense, as with the diagonalization of matrices or the representation of germs of singularities by low degree polynomials. This would seem to be a fundamental problem of topological field theory: to find the canonical elementary field theory that represents a class of more elaborate ones.

A form of topological equivalence that is less automatic than those implied by homology is when the result of a transformation a space is to produce a space of lower dimension, such as a deformation retraction onto a subspace. Indeed, in some of the elementary cases of field sources, it is necessary to perform such a retraction – usually a projection – in order for the source distribution to be defined by a cocycle. In such a case, since the homotopy type of the space does not change, and thus, the homology type remains the same, as well, one is essentially eliminating homotopically and homologically superfluous non-compact dimensions by such a transformation.

Some common examples of extended source distributions that one encounters in elementary field theory are lines, planes, solid tubes, and balls, all of which are contractible to a point. However, one should remember to also perform the same transformation on the rest of the space $M$, as well. In particular, one needs to transform $\mathsf{G}$ and its cocycles.

For instance, if the source is a line in $\mathsf{R}^3$ (say, the $z$-axis) and the field it produces is a radial vector field, so its total flux cocycle would be a 2-form, then the projection of $\mathsf{R}^3$ onto the $xy$-plane is a deformation retraction (in a trivial way) that takes the source line to a point and all cylinders that are coaxial with the source line to concentric circles about the origin. Note that it is essential that the field produced by the line source be projectable onto the $xy$-plane, which is really a statement about its symmetry, namely, its radial projection on the $z$-axis must vanish. Hence, a line source and cylindrically symmetric field in $\mathsf{R}^3$ are equivalent to a point source in the plane with a radially symmetric field. The advantage of the latter representation is that by eliminating the noncompact dimension, we have reduced our source to a cycle (in dimension zero) and our total flux functional to a cocycle (in dimension one). Indeed, before the reduction the total charge on the line and total flux through a coaxial cylinder diverged.

When one has a plane source in $\mathsf{R}^3$ (say, $yz$-plane), one can contract it to the origin by a projection onto the $x$-axis. As long as the vector field it produces is projectable – i.e., it is parallel to the $x$-axis at every point – this source/field configuration is equivalent to a point source at the origin of the real line and a vector field on it.

We illustrate the equivalent configurations for line and plane sources in $\mathsf{R}^3$ in Figure 4, in which the field is denoted by $\mathbf{E}$.

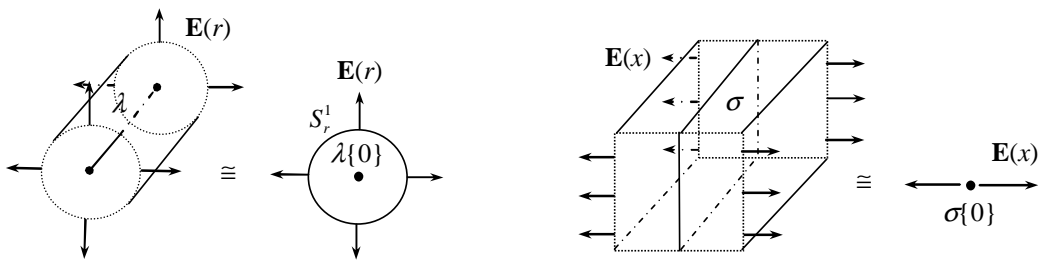

Figure 4. Equivalent field/source configurations for line and plane sources in $\mathsf{R}^3$.

## 4  Physical examples

### 4.1  Static electric fields

Of course, this class of fields includes the canonical example that motivated a lot of the abstractions that were made above, namely, the central-force vector field on $\mathsf{R}^3$:

$$\mathbf{E}(r, \phi, \theta) = \frac{1}{4\pi} \frac{Q}{r^2} \partial_r,  \tag{4.1}$$

in which $Q$ is a real number that we assume to represent a source charge or mass, $r$ is the Euclidian distance of the chosen point $(r, \phi, \theta)$ from the origin $[= (x^2 + y^2 + z^2)^{1/2}]$, and $\partial_r = \partial/\partial r$ is the radial unit vector field.

The first thing to observe is that the field is not defined at the origin, so the carrier of our singularity complex $\mathsf{S}$ is simply $\{0\}$ and our generic point set $\mathsf{G}$ is $\mathsf{R}^3 - \{0\}$. Of course, $\mathsf{S}$ has non-trivial de Rham cohomology only in dimension zero, namely, $H^0(\mathsf{S}; \mathsf{R}) = \mathsf{R}$; we shall denote the 0-cocycle that generates $H^0(\mathsf{S}; \mathsf{R})$ by $\delta^0$, and the general element of $H^0(\mathsf{S}; \mathsf{R})$ then takes the form $Q\delta^0$, where $Q \in \mathsf{R}$. As a linear functional on 0-cycles, $\delta^0$ takes the basic 0-cycle $\delta_0$, which takes 0 to $\{0\}$, to 1.

Since $\mathsf{G}$ has $S^2$ as a deformation retract, it will have non-trivial de Rham cohomology in dimension two, namely, $H^2(\mathsf{G}; \mathsf{R}) = \mathsf{R}$. We use the volume element on the unit sphere:

$$\varepsilon_2 = \sin \theta \; d\phi \wedge d\theta,  \tag{4.2}$$

as a typical representative closed 2-form for the generating de Rham cohomology class in $H^2(\mathsf{G}; \mathsf{R})$; i.e., a basis vector for the vector space. The associated linear functional on 2-cycles takes a 2-cycle to its volume when one uses $\varepsilon_2$:

$$\delta^2[z_2] = \int_{z_2} \varepsilon_2.  \tag{4.3}$$

In order to define the source/field duality isomorphism between $H^0(\mathsf{S}; \mathsf{R})$ and $H^2(\mathsf{G}; \mathsf{R})$, we could simply associate $\delta^0$ with $\delta^2$ and extend by linearity, but we prefer to induce the isomorphism by way of a connecting 3-chain in $C_3(\mathsf{R}^3; \mathsf{R})$.

Let $B(R)$ be the closed ball of radius $R$ about the origin. It can be represented as a differentiable singular cubic 3-simplex in $\mathsf{R}^3$ whose boundary is a 2-cycle in $H^2(\mathsf{G}; \mathsf{R})$, and which is contractible to the 0-cycle $\delta_0$. Hence, $B(R)$ will serve as a connecting 3-chain between the sphere of radius $R$ centered at $\{0\}$ and $\{0\}$ itself. Next, we need to show how that will imply the association of the total flux 2-cocycle $\Phi_\mathbf{E} \in H^2(\mathsf{G}; \mathsf{R})$ with the total charge 0-cycle $Q\delta^0 \in H^0(\mathsf{S}; \mathsf{R})$.

In order to compute the total flux of $\mathbf{E}$ over $\partial B(R)$, we first specify the volume element on $\mathsf{R}^3 - \{0\}$ in spherical coordinates $(r, \phi, \theta)$:

$$\varepsilon_3 = r^2 \sin \theta \; dr \wedge d\phi \wedge d\theta,  \tag{4.4}$$

so we obtain, from Poincaré duality:

$$\#\mathbf{E} = Q \sin \theta \; d\phi \wedge d\theta,  \tag{4.5}$$

and the total flux of **E** over the boundary sphere becomes:

$$\Phi_E[\partial B(R)] = \int_{\partial B(R)} \#\mathbf{E} = Q. \qquad (4.6)$$

Since we also have:

$$Q\delta^0(\delta_0) = Q, \qquad (4.7)$$

we clearly have the desired equality of total flux through the 2-sphere with the total charge at the source point.

When we take exterior derivative of #**E**, we get zero:

$$d\#\mathbf{E} = Q \cos\theta \, d\theta \wedge d\phi \wedge d\theta = 0 \, . \qquad (4.8)$$

Hence, #**E** is a closed 2-form on $\mathsf{G}$; i.e., a de Rham 2-cocycle. One also has, as a consequence:

$$\delta\mathbf{E} = 0 \, , \qquad (4.9)$$

which says that **E** is a de Rham homology 1-cycle.

Actually, equation (4.9) is not sufficient to define the vector **E**, since Helmholtz's theorem, which is really a corollary to the Hodge decomposition theorem, says that any vector field on a Riemannian manifold can be decomposed into a solenoidal part, which would satisfy (4.9), and an irrotational part. In order to make this last condition precise, one must assume that the spatial manifold – $\mathsf{R}^3$, in this case – is given a Riemannian metric $g$. Although relativistic electrodynamics would usually regard such a construction a projected artifact of the assumption of a Lorentzian structure on the spacetime manifold $M$, if one takes the "pre-metric" view of electromagnetism, which we discuss below, the spatial Riemannian structure is the real part of a complex orthogonal structure on $\Lambda^2(M)$ that is due to the electromagnetic constitutive properties of spacetime.

Such a metric gives one an isomorphism of $T(\mathsf{R}^3)$ with $T^*(\mathsf{R}^3)$, which can also be regarded as an isomorphism of $\Lambda_1(\mathsf{R}^3)$ with $\Lambda^1(\mathsf{R}^3)$ that takes the vector field **E** to the 1-form

$$E = i_{\mathbf{E}} g, \qquad (4.10)$$

which is assumed to satisfy:

$$dE = 0 \, . \qquad (4.11)$$

Hence, one assumes that the isomorphism that $g$ defines also takes the de Rham 1-cycle **E** to the de Rham 1-cocycle $E$.

When one considers static electric fields that are due to line sources, plane sources, and solid sources, one must consider that often the sources, as defined, are not representable as chains, much less cycles. For instance, a line in $\mathsf{R}^n$ or even an open line segment of finite length is not a 1-chain, and a plane or an open plane segment of finite area is not a 2-chain. Closing the finite regions with endpoints or boundaries would make the regions expressible as chains, but not cycles. Hence, we must look for equivalent field/source configurations for which the field is a cocycle and the source is a cycle; indeed, this is really what one is taught to do at the elementary level.

For the line source, one uses the cylindrical symmetry of the field/source configuration to define the predictable deformation retraction of $\mathsf{R}^3$ to $\mathsf{R}^2$ by projecting onto the $xy$-plane. (Of course, we assume that the line source is along the $z$-axis.) If we call the projection $r$ and refer to the inclusion of $\mathsf{R}^2$ in $\mathsf{R}^3$ as $(x, y, 0)$ by $i$ then the homotopy that takes the identity map to $i \cdot r$ is contraction along the $z$-axis: $(x, y, z) \mapsto (x, y, sz)$, $s \in [0, 1]$. The line source then contracts to a point source $\lambda\{0\}$ at the origin – i.e., a 0-cycle – and the vector field $\mathbf{E}$ projects to its restriction on $\mathsf{R}^2 - \{0\}$. In these expressions, $\lambda$ is the linear charge density and we had to assume that $\mathbf{E}$ had the radial symmetry about the $z$-axis that made it projectable onto the $xy$-plane.

Hence, the equivalent field source configuration is a point source of charge $\lambda$ at the origin and a vector field $\mathbf{E}$ that is defined on $\mathsf{G} = \mathsf{R}^2 - \{0\}$, so the situation is similar to the point charge in $\mathsf{R}^3$, except for the dimension. The total charge 0-cocycle $Q = \lambda\delta^0$ and the total flux 1-cycle $\Phi_{\mathbf{E}}$ associated with $\mathbf{E}$ is represented by the closed 1-form:

$$\#\mathbf{E} = \frac{\lambda}{2\pi}d\theta \, . \tag{4.12}$$

Any circle of radius $r$ that is centered at the origin will serve as a partial 2-boundary $z_2$ in $\mathsf{G}$, such that the closed disc of radius $r$ that it bounds in $\mathsf{R}^3$ is the connecting 3-chain in $\mathsf{R}^3$ that is contractible to the source 0-cycle. Equating the value of $\Phi_{\mathbf{E}}$ when evaluated on $z_2$ with the value of $Q$ when evaluated on $\delta_0$ gives a $1/r$ dependency for the resulting $\mathbf{E}$ field:

$$\mathbf{E}(r, \theta) = \frac{\lambda}{2\pi r}\frac{\partial}{\partial r} \, . \tag{4.13}$$

In the case of the plane source (we use the $yz$-plane for specificity) of constant surface charge density $\sigma$, one uses symmetry to project $r: \mathsf{R}^3 \to \mathsf{R}$, $(x, y, z) \mapsto (x, 0, 0)$, so the homotopy from the identity on $\mathsf{R}^3$ to $i \cdot r$ takes $(x, y, z)$ to $(x, sy, sz)$, $s \in [0, 1]$.

The equivalent field/source configuration is then a source of charge $\sigma$ at $\{0\}$ on $\mathsf{R}$ and an $\mathbf{E}$ field that is defined on $\mathsf{G} = \mathsf{R} - \{0\}$. Hence, the total charge 0-cocycle is simply $\sigma\delta^0$ and the total flux 1-cycle associated with $\mathbf{E}$ is represented by the closed 0-form:

$$\#\mathbf{E} = \sigma \, . \tag{4.14}$$

The partial boundaries in $\mathsf{R} - \{0\}$ that one uses are "0-spheres of radius $r$ centered at $\{0\}$," namely the pairs of points at $\pm r$. The connecting 1-chain in $\mathsf{R}$ is the closed interval $[-r, +r]$. Since there are two points in the partial boundary, the resulting $\mathbf{E}$ field is:

$$\mathbf{E}(x) = \pm \frac{\sigma}{2} \frac{\partial}{\partial x}, \qquad \text{sgn}(\mathbf{E}) = \begin{cases} + & x > 0 \\ - & x < 0 \end{cases}, \tag{4.15}$$

Note that in this case, the singularity at {0} takes the form of a finite jump discontinuity, instead of a pole.

### 4.2 Static magnetic fields

There are some fundamental differences between static magnetic fields and static electric fields:

*a.* Although a transient magnetic field can have an "open circuit" for a circuit, a stable magnetic field must have a constant current for a source, which can only exist in a conducting loop; i.e., the source complex S must be a 1-cycle $I_1 = I\delta_1$, in which we are representing the current in the loop by the real number $I$ and $\delta_1 \in H_1(\mathsf{S}; \mathsf{R})$, which is the "fundamental 1-cycle" that represents the orientation of the loop in singular homology, just as a choice of non-zero tangent vector field would in de Rham homology.

*b.* The magnetic field strength is most directly represented as a covector field $B$; i.e., a 1-cochain $B \in C^1(\mathsf{G}; \mathsf{R})$, instead of a vector field. This is due to the fact that the form of Stokes's theorem that one uses is the two-dimensional Ampèrian form, so the fundamental integral involves a 1-form on G and a 1-cycle $z_1$ in G, namely:

$$<B, z_1> = \int_{z_1} B \,, \tag{4.16}$$

which one calls the *magnetomotive force* [8] *(mmf) around* $z_1$, instead of the total flux of a vector field through a surface.

*c.* Since $B$ is a 1-form, the conventional equation that couples $B$ with $I$, namely, $dB = 4\pi \#I$, involves $d$ directly, not $\delta$, and also suggests that $B$ does not define a cohomology class when $I \neq 0$. However, we are confronting a variation of the situation in the previous section, because when one (naively) applies Ampère's law to $B$ and $I$ when $z_1 = \partial D_2$, where the 2-chain $D_2$ is a disc of radius $r$ in $\mathsf{R}^3$ that $\delta_1$ intersects transversely:

$$<B, \partial D_2> = <dB, D_2> \ = <\#I, D_2> = <\delta^1, I_1> = I \,, \tag{4.17}$$

one finds that the resulting magnetic field (when the source current loop has a sufficiently large radius compared to $r$) has the form:

$$B(r) = \frac{I}{2\pi r} ds = \frac{I}{2\pi} d\theta \,, \tag{4.18}$$

in which $ds = r \, d\theta$ is the arc-length 1-form. This gives:

$$dB = 0. \tag{4.19}$$

---

[8] The use of the word "force" is actually a misnomer, since the units of the quantity are work done per unit magnetic "charge."

Hence, since $B(r)$ is not defined at the center of $D_2$, Stokes's theorem is inapplicable, and the association of $<B, \partial D_2>$ with $I$ in (4.17) is by definition, not by consequence. Basically, we are back to the previous situation for the electric field of a line source, except that the 1-cocycle $B$ is tangent to the boundary of $D_2$, not normal, and the source is the 0-cycle that is defined by intersection of $\delta_1$ with $D_2$, not a contraction of $\delta_1$.

### 4.3 Dynamic fields and inductive couplings

In this section, we are really addressing a situation for which the machinery that we have been developing for topological field theories is no longer applicable, because the source of one field does not have a cycle for its carrier, but a chain $c_k$ with boundary. In particular, the source of one field $\alpha_{k-1} \in \Lambda^{k-1}(\partial c_k)$ defined on $\partial c_k$ will be a function of the other field $\beta_k \in \Lambda^k(c_k)$ that is defined on $c_k$. The level at which they are coupled is their evaluations:

$$<\alpha_{k-1}, \partial c_k> = F(<\beta_k, c_k>). \tag{4.20}$$

Of course, in this case Stokes's theorem applies, so we can also say that:

$$<d\alpha_{k-1}, c_k> = F(<\beta_k, c_k>). \tag{4.21}$$

and if this is true for all $c_k$ then we can say that:

$$d\alpha_{k-1} = F'(\beta_k) \tag{4.22}$$

for some other function $F'$. Hence, we see that the $k-1$-form $\alpha_{k-1}$ is generally not closed, and therefore will not define a de Rham cohomology class, but only a $k-1$-cochain.

Of course, the motivating examples for the present situation are defined by electromagnetic induction. Faraday's law couples the electromotive force around a (bounding) 1-cycle $z_1 = \partial c_2$ in $\mathsf{R}^3$ to the time derivative of the total magnetic flux through $c_2$:

$$<E, \partial c_2> = <dE, c_2> = -\frac{d}{dt}<B, c_2>. \tag{4.23}$$

Conversely, Maxwell's law of induction couples the magnetomotive force around $z_1$ to the time derivative of the total electric flux through $c_2$:

$$<B, \partial c_2> = <dB, c_2> = +\frac{d}{dt}<E, c_2>. \tag{4.24}$$

Note the differing signs between (4.23) and (4.24); without this, there would be no electromagnetic waves.

### 4.4 Topological electromagnetism

So far, we have considered three-or-less-dimensional spaces in which static electric and magnetic fields were defined, and then suggested that topological methods broke down for time-varying fields due to the non-conservation of energy in the loops; i.e., the fact that they bounded 2-chains through which time-varying field fluxes were defined. This is not entirely accurate, since the restriction to dimension three or less is a sign of incompleteness in the model, as far as relativistic electrodynamics is concerned.

If we go to the four-dimensional spacetime $M$ that relativity suggests then we can reformulate the problem of source/field duality quite simply. Although the usual relativistic formulation of electromagnetism uses a Lorentzian structure on the tangent bundle $T(M)$, since we are discussing topological electromagnetism, we shall regard the reduction of the bundle of linear frames to the bundle of Lorentzian frames as a consequence of the electromagnetic structure of spacetime, not a prerequisite for it.

This is in the spirit of pre-metric electromagnetism (see [11-13]), in which a more fundamental role is played by the introduction of a unit-volume element $\varepsilon \in \Lambda^4(M)$ on $T(M)$, which we naturally assume to be orientable, and a linear electromagnetic constitutive law $\kappa: \Lambda^2(M) \to \Lambda_2(M)$. This represents a field of isomorphisms from 2-forms at each point of $M$ to bivectors at the same point. When one composes these isomorphisms with the Poincaré duality isomorphism #: $\Lambda_2(M) \to \Lambda^2(M)$ that comes from $\varepsilon$, one gets an automorphism of $\Lambda_2(M)$ that, by hypothesis, differs from an automorphism * of $\Lambda_2(M)$ that has the property $*^2 = -I$ only by a non-zero scalar multiple at each point.

The latter automorphism basically gives us what the Hodge duality isomorphism would have given us if we had introduced a Lorentzian structure, but only for 2-forms. It also defines an almost-complex structure on $\Lambda^2(M)$ that makes each fiber non-canonically C-isomorphic to $C^3$. The group of linear transformations of fibers of $\Lambda^2(M)$ that preserve both the linear electromagnetic structure and the unit-volume element – hence, the almost-complex structure – is isomorphic to $GL(3; C)$. The unit-volume element on $T(M)$ can be used to define a unit-volume element on $\Lambda^2(M)$, which then allows one to reduce the group to $SL(3; C)$. One can then use $\varepsilon$ and * to define a complex orthogonal structure on $\Lambda^2(M)$, which then allows a reduction to $SO(3; C)$, which is isomorphic to the identity component of the Lorentz group, viz., the proper orthochronous Lorentz group. Hence, oriented, time-oriented Lorentzian frames in the tangent spaces to $M$ are in one-to-one correspondence with oriented complex orthogonal 3-frames in the fibers of $\Lambda^2(M)$. This is basically the point at which pre-metric electromagnetism rejoins the metric version.

Hence, we assume that our four-dimensional spacetime $M$ has a unit-volume element $\varepsilon$ on its tangent bundle and an almost-complex structure on its bundle of 2-forms. As far as topological things are concerned, it is more directly useful to define isomorphisms on the exterior algebras over $T(M)$ and $T^*(M)$ themselves, rather than start with isomorphisms on $T(M)$ and $T^*(M)$ and then induce isomorphisms in the tensor algebra, as is usually done in metric differential geometry.

The kind of differential geometry that deals with the bundle of 2-forms as fundamental, rather the tangent bundle, is called *line geometry* [14], which is a special case of projective differential geometry. This is because, under the Plücker embedding of $RP^5$ in $P\Lambda_2(R^4)$, a line in $RP^5$ (which is also a 2-plane in $R^6$) maps to an equivalence class of decomposable bivectors that differ by a non-zero real scalar. The set of all these decomposable bivectors defines a quadric hypersurface in $P\Lambda_2(\mathbb{R}^4)$ that is called the *Klein quadric*. However, considering the central role that is played by the almost-complex structure *, one should also look at the complex version of this embedding, which takes a *point* in $\mathbb{CP}^2$ to an equivalence class of non-zero decomposable bivectors that differ by a non-zero *complex* scalar, which is just the projective equivalence of $\mathbb{CP}^2$ with complex lines through the origin in $\Lambda_2(\mathbb{R}^4)$, when it is given a complex structure.

As far as the topological considerations are concerned, we first formulate the pre-metric Maxwell equations as:

$$dF = 0, \qquad \delta\mathbf{h} = \mathbf{J}, \qquad \mathbf{h} = \kappa(F). \tag{4.25}$$

In these equations, $F \in \Lambda^2(M)$ is the Minkowski 2-form of electromagnetic field strengths, $\mathbf{h} \in \Lambda_2(M)$ is the bivector field of electromagnetic excitations, and $\mathbf{J}$ is the source current for the field $\mathbf{h}$, and consequently $F$.

As a consequence of the field equations, $\mathbf{J}$ has the property that:

$$\delta\mathbf{J} = 0. \tag{4.26}$$

Hence, we see that $\mathbf{J}$ defines a de Rham 1-cycle.

Since one of the Maxwell equations says explicitly that $F$ is a de Rham 2-cocycle, we see that we are indeed looking at the topological field theoretic situation that we have been discussing up till now.

### 4.5 Fluid flow

Since an incompressible fluid in steady-state flow is governed by the same equations as a static electric field – i.e., the vanishing of a divergence – we shall only summarize the details of the analogy in this section.

The fundamental field that one is concerned with is the fluid flow velocity vector field $\mathbf{v}$. Since we are assuming steady-state flow, it can be defined as a vector field with compact support in $\mathbb{R}^3$. To say that the flow is incompressible is to say that the volume element $\varepsilon$ is constant along the flow of $\mathbf{v}$. Hence, the Lie derivative of $\varepsilon$ with respect to $\mathbf{v}$ vanishes:

$$0 = L_\mathbf{v}\varepsilon = di_\mathbf{v}\varepsilon = d\#\mathbf{v} = \#\delta\mathbf{v}. \tag{4.27}$$

Since # is an isomorphism, $L_\mathbf{v}\varepsilon = 0$ iff $\delta\mathbf{v} = 0$.

Hence, the flow velocity vector field behaves just like $\mathbf{E}$ for our present purposes. The curves that have $\mathbf{v}$ for their tangent vectors are called *streamlines*, as opposed to the field lines that one associates with $\mathbf{E}$.

The analogues of charge distributions that represent the sources (and sinks) of $\mathbf{v}$ are regions of space, which will be the $\mathbf{S}$ of this scenario, in which fluid mass is being introduced or taken away from the region in which $\mathbf{v}$ is defined, which is what $\mathbf{G}$ amounts to. Usually, this takes the form of a channel or tank in practical fluid flow models. A point source/sink is really an approximation to a tube of small diameter when compared to the dimensions of the tank or channel.

One also has a close analogy between the theory of steady-state fluid vortices and static magnetic fields. The flow velocity vector field **v**, or rather, its covelocity 1-form $v = i_\mathbf{v}\delta$, plays the role of the magnetic vector potential field $A$ and its vorticity 2-form $\Omega = dA$ plays the role of $B$. (Here, $\delta$ refers to the Euclidian scalar product on $\mathsf{R}^3$.) The analogue of an electric source current along a curve is called a *vortex filament*. As with magnetic fields, one often formulates the field theory as two-dimensional with a point source.

One can even extend the analogy between fluid flow and electromagnetism to an analogy between relativistic hydrodynamics (see [**15, 16**]) and Maxwell electrodynamics. One starts with either the relativistic fluid flow covelocity 1-form $u$ or the energy-momentum 1-form $p$ as the analogue of the electromagnetic potential 1-form, which implies that the timelike component of covelocity or energy-momentum is analogous to the electric charge density and the spacelike velocity or momentum 1-form is analogous to the magnetic vector potential [9]. The 2-form $du$ is referred to as the *kinematical vorticity* and the 2-form $dp$ as the *dynamical vorticity*. In a 1+3 decomposition of these 2-forms, the electric part of the 2-form describes a form of linear acceleration or force, respectively, and the magnetic part describes either the angular velocity or the angular momentum, respectively.

### 4.6 Weak-field gravitation

For centuries, it has been accepted that there is a formal analogy between the theory of static weak-field (i.e., Newtonian) gravitational fields and static electric fields, although the difference between the character of the respective forces (i.e., how the signs of the charges/masses determine whether the force is one of attraction or repulsion) is perplexing considering that nobody has experimentally isolated any negative masses to establish whether they actually attract or repel the more conventional positive masses.

The basis for the analogy is that the gravitational acceleration vector field **g** obeys equations of the same form as the field equations for the electric field strength vector field **E** when the source charge distribution $\rho$ is static, namely:

$$dg = 0, \qquad \delta\mathbf{g} = \rho\,. \tag{4.28}$$

In these equations, $g = i_\mathbf{g}\delta$ is the metric-dual gravitational co-acceleration 1-form, relative to the Euclidian scalar product $\delta$ on $\mathsf{R}^3$, and $\rho$ is the mass density distribution.

As a result of this analogy, all of the discussion that related to electrostatics as a topological field theory has a direct analogue in the static gravitational context. In particular, when the source distribution is a point mass $m$ the field **g** is defined only on $\mathsf{G} = \mathsf{R}^3 - \{0\}$ and one has that $\delta\mathbf{g} = 0$, so **g** defines a de Rham 1-cycle on $\mathsf{G}$ and $m$ defines a singular 0-cycle. The Poincaré dual 2-form #**g** then defines a mass flux density whose integral over a 2-cycle will equal $m$, despite the vanishing of the divergence of **g**.

---

[9] This suggests an interesting analogy between gauge invariance and the relativity of velocity. In effect, the kinematical or dynamical vorticity 2-form takes on the physically fundamental role and a choice of covelocity or energy-momentum 1-form, resp., involves a choice of "gauge", which would be an irrotational covelocity or energy-momentum 1-form, resp.

An intriguing possibility that has been investigated almost since the earliest days of Maxwellian electrodynamics is that the analogy between weak-field gravitation and electrostatics actually extends to an analogy between dynamic gravitational fields and Maxwellian electrodynamics. The key to extending the analogy would the establishment of a "gravito-magnetic" field $\mathbf{h}$ that would be associated with a mass current vector field $\mathbf{p} = \rho\mathbf{v}$ – i.e., a momentum density vector field – in the same way that the magnetic field $\mathbf{B}$ (or $\mathbf{H}$) is associated with an electric charge current vector field. The main obstacle to establishing "gravito-electromagnetism" (see [**17**] for a survey of the issues) is the fact that even for the much more powerful electromagnetic interaction the forces that derive from the existence of a magnetic field induced by the motion of electric charges are generally much weaker than the electrostatic forces, and the gravitational analogue of the electrostatic force is quite weak to begin with. A compelling direction of investigation for exhibiting this force as something that occurs in Nature is the possibility that the gravito-magnetic field of a spiral galaxy might account for the anomalous behavior of its angular velocity versus distance from the center curve without the necessity of introducing a new form of matter – viz., dark matter – that has not been observed in any other context [10]. Similarly, it might play an important role in the dynamics of the Big Bang. Indeed, it seems clear that a force as weak as the gravito-magnetic force would only become noticeable for enormous mass currents of astrophysical proportions.

## 5 Discussion

One of the differences between the role of singularities in physics and their role in mathematics is that mathematics has the freedom to postulate the character of its pathologies, whereas physics is ultimately at the mercy of the limits of experimental resolution to document them. Hence, since the interior of a singular region of physical space always has some element of enigma to it, one needs to have a compelling reason to extend one's theoretical model beyond the limits of experiment.

Commonly, one reasons by analogy, but, it often helps to also have some sort of "eversion" principle that allows one to associate the phenomena that transpire inside the region with the phenomena outside of it. For instance, one models the interior of the Earth on the basis of the way that it refracts, reflects, and disperses seismic waves that originate on the surface. Field/source duality is an attempt at associating at least the topological information inside of a singular region of space with corresponding information that is manifest in the exterior region.

Although it may seem restricted to only those physical fields that have sources consisting of points at which the field is undefined or discontinuous in one of its derivatives, this is not as severe a restriction as it may sound. Indeed, one suspects that this is the nature of the most fundamental level of field sources.

There is more that must be learned about the purely mathematical problem of the structure of manifolds that admit such a duality. From the examples given above, it is certainly present whenever the generators of the homology or cohomology in some dimension were created by the simple act of removing a set of open disjoint balls of that dimension. One naturally inquires whether this is the only way that the connecting chains come about.

---

[10] This possibility was suggested by Lowell Cummings in unpublished work.

In a subsequent study the further consequences of regarding field sources as being more specifically topological obstructions to the integrability of the field equations will be investigated.